\documentclass{PoS}

\usepackage{graphicx}
\usepackage{natbib}
\usepackage[figuresleft]{rotating}
\newcommand{\HI}{H{\,\small I}}
\newcommand{\ltsima} {$\; \buildrel < \over \sim \;$}
\newcommand{\gtsima} {$\; \buildrel > \over \sim \;$}
\newcommand{\lta} {\lower.5ex\hbox{\ltsima}}
\newcommand{\gta} {\lower.5ex\hbox{\gtsima}}

\newcommand{\FRI}{FR{-\small I}}
\newcommand{\FRII}{FR{-\small II}}

\title{HI in radio galaxies: prospects for upcoming wide-field surveys}

\ShortTitle{HI in radio galaxies}

\author{\speaker{Bjorn Emonts}\\
        CSIRO Australia Telescope National Facility, PO Box 76, Epping NSW, 1710, Australia\\
        E-mail: \email{bjorn.emonts@csiro.au}}

\author{Raffaella Morganti, Christian Struve\\
        Netherlands Foundation for Research in Astronomy, Postbus 2, 7990 AA Dwingeloo, NL\\
        Kapteyn Astronomical Institute, University of Groningen, P.O. Box 800, 9700 AV Groningen, NL}

\abstract{We present results of an ongoing systematic study of the large-scale properties of neutral hydrogen (\HI) gas in nearby radio galaxies. Our main goal is to investigate the importance of gas-rich galaxy mergers and interactions among radio-loud AGN. From an \HI\ study of a complete sample of classical low-power radio galaxies we find that the host galaxies of extended Fanaroff $\&$ Riley type I radio sources are generally \HI\ poor ($\lesssim 10^8 M_{\odot}$) and show no indications for gas-rich galaxy mergers or ongoing gas-rich interactions. In contrast, the host galaxies of a significant fraction of low-power compact radio sources contain enormous discs/rings of \HI\ gas (with sizes up to 190 kpc and masses up to $2 \times 10^{10} M_{\odot}$). This segregation in \HI\ mass with radio source size likely indicates that these compact radio sources are either confined by large amounts of gas in the central region, or that their fueling is inefficient and different from the fueling process of classical \FRI\ radio sources. To a first order, the overall \HI\ properties of our complete sample (detection rate, mass and morphology) appear similar to those of radio-quiet early-type galaxies. If confirmed by better statistics, this would imply that low-power radio-AGN activity may be a short phase that occurs at some point during the lifetime of many early-type galaxies. We discuss how upcoming \HI\ surveys (e.g. with ASKAP and Apertif) are essential for studying in a statistical way the the connection between the presence and morphology of a radio-loud AGN and the properties of the cold \HI\ gas associated with its host galaxy.}

\FullConference{Panoramic Radio Astronomy: Wide-field 1-2 GHz research on galaxy evolution - PRA2009\\
		 June 02 - 05 2009\\
		 Groningen, the Netherlands}

\begin{document}

\section{Introduction}
Active Galactic Nuclei (AGN) are believed to be triggered when gas and matter is deposited onto a super-massive black hole in the center of the host galaxy. For this to happen, the gas needs to lose sufficient angular momentum to be transported deep in the potential well of the galaxy, until it eventually fuels the AGN. Galaxy mergers and interactions have often been invoked as a possible triggering mechanism for AGN activity. During a merger/interaction, perturbations in the galactic potential may stir the gas in a galaxy and enhance cloud-cloud collisions . This may result in the fueling of the central black hole, the exact timing of which could be related to secondary processes, such as bar formation or the merging of individual black holes of the progenitor galaxies. 

We are in the process of systematically investigating the merger/interaction history of well-defined samples of various types of nearby radio galaxies through observations of the large-scale neutral hydrogen (\HI) gas. While in a galaxy merger or interaction part of the gas is driven into the central region, another part may be expelled in large structures \citep[tidal-tails, bridges, shells, etc., e.g.][]{bar02}. If the environment is not too hostile, part of these gaseous structures (which often have a too low surface density for violent star formation to occur) remain bound to the host galaxy as relic signs of the galaxies' violent past. \HI\ observations, in particular when combined with optical information about the stellar content, are therefore an excellent tool to trace and date galaxy mergers or interactions over relatively long time-scales. \HI\ observations of nearby radio galaxies have the additional advantage that absorption studies can be done against the radio continuum, allowing in many cases to detect \HI\ to much lower column densities than what can be traced in emission. Besides mergers/interactions, other AGN feeding mechanisms (such as the accretion of hot gas from the circum-galactic medium) are often ascribed to certain types of radio galaxies \citep[e.g.][]{all06,balma08}. Indications for such processes may also be reflected in the in the \HI\ content (or lack thereof) of nearby radio galaxies.

In this paper we present the \HI\ results of a complete sample of nearby low-power radio galaxies. This complete sample consists of 22 radio galaxies from the B2-catalogue ($F_{\rm 408MHz} \gtrsim 0.2$ Jy) up to a redshift of {\sl z} = 0.041, plus radio galaxy NGC~3894 (which has properties similar to the B2 sources, but falls just outside the declination completeness limit of the B2 sample). Sources in dense clusters and BL Lac objects are excluded from our sample. Our complete sample contains both classical \FRI\ as well as compact radio sources, all with a radio power 22.0 $<$ Log ($P_{\rm 1.4\ GHz}$) $<$ 24.8 (and with no bias in $P_{\rm 1.4\ GHz}$ between the compact and extended sources). Observations were done with the Very Large Array in C-configuration and the Westerbork Synthesis Radio Telescope. 

The next Section gives an overview of the results of this \HI\ study of low-power radio galaxies \citep[to be published in an upcoming paper; ][]{emo10} as well as a preview of our ongoing work on large-scale \HI\ in more powerful radio galaxies. In Sect. \ref{sec:askap} we discuss how upcoming \HI\ all sky-surveys with ASKAP and Apertif form a crucial next step in this research.

\section{Results}
\label{sec:results}

Our detection rate of \HI\ emission associated with the host galaxies of nearby low-power radio sources is 29$\%$ (with a detection limit of $\sim 10^8 M_{\odot}$). Interestingly, {\sl none} of the 14 early-type host galaxies of extended \FRI\ radio sources in our sample contain $> 10^{8} M_{\odot}$ of \HI\ emission-line gas. For two \FRI\ sources in our sample we observe clouds of \HI\ gas with a total mass of $\sim 7 \times 10^{7} M_{\odot}$ (i.e. close to the detection limit of most of our sample sources), while one object in our sample with $M_{\rm HI} = 2.3 \times 10^8 M_{\odot}$ is a very rare spiral-hosted \FRI\ radio source \citep{emo09}. Of the remaining non-detected \FRI\ sources, only one shows tentative evidence for \HI\ absorption. This indicates that, while small amounts of \HI\ gas may be present in some \FRI\ radio galaxies, these systems are generally \HI\ poor.

\begin{figure}
\centering
\includegraphics[width=\textwidth]{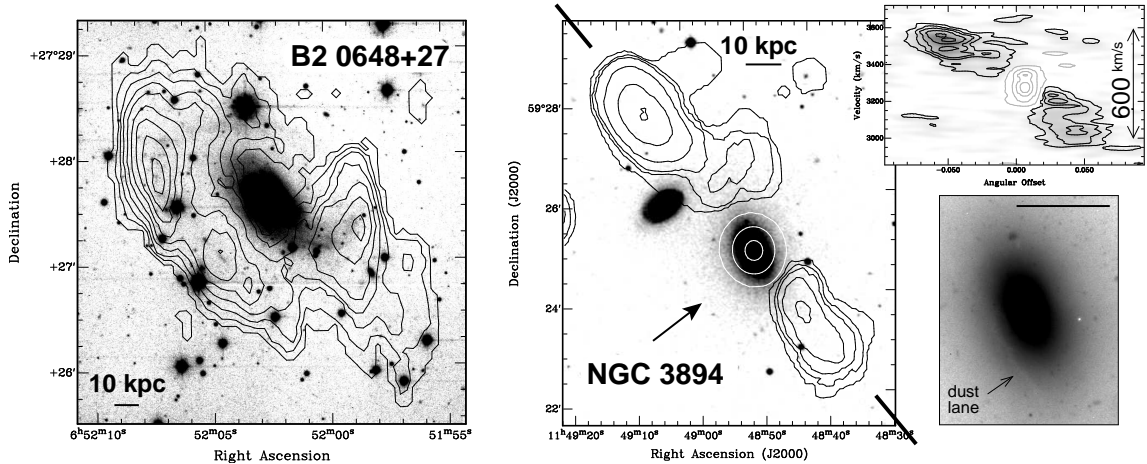}
\caption{{\sl Left:} B2~0648+27 -- disc properties: $M_{\rm HI} = 8.5 \times 10^{9} M_{\odot}$ and diameter 190 kpc. {\sl Right:} NGC~3894 -- disc properties: $M_{\rm HI} = 2.2 \times 10^{9} M_{\odot}$ and diameter 105 kpc. More details are given in \citet{emo10}.}
\label{fig:HIdiscs}
\end{figure}

On the other hand, the four most compact radio sources in our sample all contain enormous \HI\ discs/rings, with \HI\ masses between $7 \times 10^{8} - 2 \times 10^{10} M_{\odot}$ and diameters of $37 - 190$ kpc. Two of these \HI\ discs are shown in Fig. \ref{fig:HIdiscs}. We classify B2~0648+27 (Fig. \ref{fig:HIdiscs} - {\sl left}) as a gas-rich post-merger system, given the distorted optical morphology of the system and the presence of a dominating post-starburst stellar population throughout the host galaxy \citep[see][]{emo06,emo08_0648}. We argue that the large-scale \HI\ ring \citep[which has a faint optical counterpart in the form of a long tail or partial stellar ring;][]{emo08_0648} formed as a result of this merger, which must have started roughly 1.5 Gyr ago \citep{emo06}. Since then, gas and stars that were expelled during the merger have had the time to fall back onto the host galaxy and settle into the regularly rotating ring that we observe today. The disc/ring around NGC~3894 has no apparent optical counterpart (Fig. \ref{fig:HIdiscs} - {\sl right}). The host galaxy has a much more regular elliptical morphology, although a very faint dust-lane crosses the host galaxy in roughly the same direction as the \HI\ disc. Since we see the \HI\ disc in NGC~3894 edge-one, there could be significant extinction within the \HI\ disc. It is possible that the \HI\ disc in NGC~3894 formed from a much older merger event, or though a steady accretion of \HI\ gas from the inter-galactic medium over time-scales of several Gyr \citep[see][for more details]{emo07,emo10}.

\begin{figure}
\centering
\includegraphics[width=0.55\textwidth]{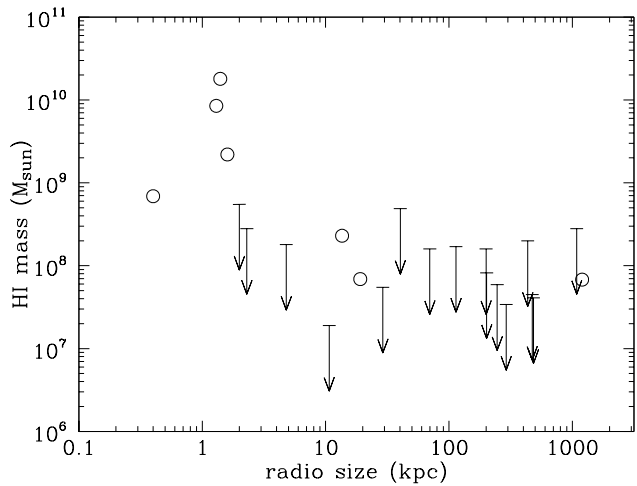}
\caption{From \citet{emo07}. Total \HI\ mass detected in emission plotted against the linear size of the radio sources. In case of non-detection a tight upper limit (3$\sigma$ across 200 km s$^{-1}$) is given.}
\label{fig:HImassplot}
\end{figure}

Figure \ref{fig:HImassplot} \citep[from][]{emo07} clarifies the striking difference in \HI\ content between the extended \FRI\ sources and the low-power compact sources in our sample \citep[note that we find no bias in \HI\ content with respect to the total power of the radio source or the absolute visual magnitude and optical morphology of the host galaxy;][]{emo10}. This segregation in \HI\ mass between compact and extended low-power radio sources suggests that there is a physical link between the properties of the radio source and the presence of large-scale \HI\ structures. The most likely explanation for this segregation is that either the compact sources are confined/frustrated by large amounts of gas that is present also in the central region, or that their fueling mechanism (likely related directly to the cold gas in these systems) is much less efficient than that of extended \FRI\ sources, which prevents them from growing into extended sources.

Although we saw that some of the regularly rotating \HI\ discs/rings may be a relic of a past merger event \citep[> Gyr old, i.e. much older than the typical life-time of radio sources;][]{par02}, across our sample of low-power radio galaxies we find no evidence from the distribution/kinematics of \HI\ gas that {\sl ongoing} gas-rich galaxy mergers or violent galaxy interaction are associated with these systems. This fact is strengthened by comparing our \HI\ results with similar recent studies of nearby early-type galaxies that were not selected on radio loudness \citep[][2009 (in prep.)]{mor06,oos07}. Because the number density of radio galaxies is low and they are relatively far away compared with nearby radio-quiet early-type galaxies, the sensitivity of the various studies is not uniform. Nevertheless -- as we will describe in detail in \citet{emo10} -- when taking the difference in detection limit into account, we find no apparent major differences in the detection rate, \HI\ mass or morphological characteristics of the \HI\ structures between our sample of low-power radio galaxies and the above mentioned samples of radio-quiet(er) early-type galaxies. For sure, across the range of \HI\ masses that we studied in this paper, there is no evidence that our radio-loud sample has a higher detection-rate or contains more tidally distorted \HI\ structures than the radio-quiet samples. If confirmed by larger samples with uniform sensitivity, this may indicate that the radio-loud phase could be just a short period that occurs at some point during the lifetime of many -- or maybe even all? -- early-type galaxies.

The general lack of observable amounts of \HI\ gas, in combination with the similarity in \HI\ content between radio-loud and radio-quiet early-type galaxies, is in agreement with the growing evidence that classical \FRI\ sources are fed through a steady quasi-spherical Bondi accretion of the hot inter-galactic medium directly onto the central black-hole \citep[e.g.][]{all06,balma08}. As we saw two paragraphs earlier, either the fueling mechanism or the evolution of the \HI-rich compact radio sources in our sample is likely fundamentally different from that of the extended \FRI\ sources and somehow related to the presence of large amounts of \HI\ gas.

The situation may be different for more powerful \FRII\ radio galaxies. From optical studies, \citet{hec86} and \citet{bau92} argued that powerful radio galaxies with strong emission-lines often show peculiar optical morphologies and emission-line features reminiscent of a gas-rich galaxy merger (a property generally not shared among low-power radio galaxies). In an ongoing study, we started to map the \HI\ content of a small but complete sample of the nearest \FRII\ radio galaxies in order to investigate whether powerful radio galaxies show a fundamental difference in \HI\ content -- and hence formation history -- compared with \FRI\ systems. In a recent publication \citep{emo08_NGC612} we show that the nearby powerful radio galaxy NGC~612 does indeed contain large amounts of \HI. While also here the bulk of the gas is distributed in a large disc (with $M_{\rm HI} = 1.8 \times 10^{9} M_{\odot}$ and diameter = 140 kpc), a faint \HI\ bridge stretches across 400 kpc toward a gas-rich companion galaxy, indicating that a collision between both systems likely occurred. \HI\ results on the complete sample of powerful \FRII\ radio galaxies, as well as a comparison with our current sample of low-power radio galaxies, will be presented in a future paper.

\section{All-sky \HI\ surveys (ASKAP/Apertif)}
\label{sec:askap}

Wide-field survey radio telescopes currently under construction, such as the Australian Square Kilometre Array Pathfinder (ASKAP) and Apertif on the Westerbork Synthesis Radio Telescope, will enable all-sky \HI\ surveys at a sensitivity and resolution that will be very similar to the observations mentioned in this paper (detection limit $M_{\rm HI} \sim 10^{8} M_{\odot}$). This will provide, for the first time, \HI\ information (both in emission and in absorption against the radio continuum) on large statistical samples and sub-samples of well-defined types of nearby radio galaxies. 

In this Section, we show the feasibility of a large project on \HI\ in radio galaxies based on all-sky survey projects that are currently in preparation for ASKAP (although similar conclusions are reached for an all-sky \HI\ survey with Apertif). Taking into account the \HI\ detection rate and selection criteria of our B2 sample of radio galaxies (presented in this paper) and considering the fact that the B2 catalogue covers an area less than 20$\%$ of that of the all-sky \HI\ survey for ASKAP (WALLABY, or Wide-field ASKAP L-band Legacy All-sky Blind surveY), we we estimate that WALLABY will result in roughly 150 \HI\ emission-line detections out of a sample of $10^3$ radio galaxies (field + cluster) with a continuum flux of $S_{\rm 1.4 GHz} \gtrsim 50$ mJy out to $z = 0.05$. We argue that this is merely a lower limit when taking into account the full range of radio-source powers that we studied in our current B2 sample, since at a typical redshift of $z = 0.03$ a low-power radio source with $P_{\rm 1.4 GHz} = 10^{22}$ W/Hz (i.e. similar to the weakest radio sources in our current B2 sample) has a flux that is an order of magnitude lower than our conservative estimate of 50 mJy (for those sources, $S_{\rm 1.4 GHz} \sim 5$ mJy, which will be easily observable with the ASKAP all-sky continuum survey EMU [Evolutionary Map of the Universe]). Note also that this estimate only includes \HI\ detected in emission; all-sky \HI\ absorption studies (e.g. the First Large Absorption Survey in \HI\ [FLASH]) will reveal numerous more detections and address additional interesting science questions. One key element in reaching the full potential of a large study on \HI\ in radio galaxies will be the requirement of sufficiently high spectral dynamic range and stable bandpass calibration in order to look for faint \HI\ detections in the vicinity of relatively strong continuum sources.

Although most of the radio sources expected to be studied with ASKAP and Apertif fall in the category of lower-power FR-I sources, many of those will be compact and a few dozen will be powerful FR-II radio sources. This is enough for studying the difference in \HI\ properties between the various types of radio galaxies (which will provide valuable information for science cases for the Square Kilometre Array [SKA], which will be able to reach much further into the Universe, where powerful radio galaxies are more numerous). In addition, an all-sky survey will provide 'for free' a complementary study of the \HI\ properties of radio-quiet early-type galaxies at the same sensitivity, which will be ideal for comparing the \HI\ properties of radio galaxies with that of their radio-quiet counterparts. As a follow-up of an extensive \HI\ survey of nearby radio galaxies, the Atacama Large Millimeter Array (ALMA) can provide crucial information at high angular resolution about the molecular gas in the very central region close to the black-hole, allowing to investigate directly the fueling mechanism of the AGN.

\end{document}